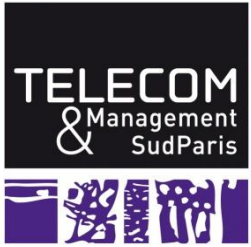
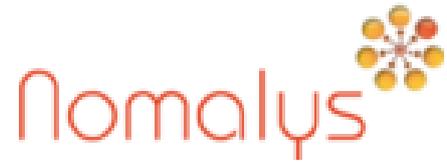

# THESE

**Spécialité** : Management des Télécoms

Présentée pour obtenir le titre :
# Manager Telecom

Par

# Djallel Bouneffouf

**Stage effectué à l'entreprise Nomalys**

Sujet :

*Proposition et déploiement d'une technique de gestion de projet dans l'entreprise NOMALYS*

# Résumé du projet

Ce projet s'insère dans le cadre du développement de CRM embarqué. Il a pour objectif le développement d'une application de gestion de la clientèle à distance, Nommé NOMALYS. Cette application permet au commercial et chefs d'entreprises de consulter leur CRM sur mobile.

Je me suis focalisé dans ce projet à étudier les techniques de gestion de projet, Cette étude m'a permis de classer différent technique de gestion existante et de proposer la technique qui correspond le plus au besoin de l'entreprise.

# Sommaire

# Tableau des figures

# Introduction

Nous regroupons sous le vocable (*Management de projet)* l'ensemble des techniques visant à structurer, assurer et optimiser le bon déroulement d'un projet.

Pour arriver à un résultat satisfaisant, les techniques de gestion de projet doivent permettre aux responsables de projet un gain de temps et d'argent.

De nos jours les projets informatiques et télécoms ne font que se diversifier et les techniques de management doivent suivre l'évolution et s'adapter à la diversification.

## Motivation

Notre approche est motivée par les constatations suivantes :
Était une startup venant de naitre, tous était à faire et s'appuyer sur des préceptes de conduite de projet était la plus essentielle pour arriver au résultat attendu dans le respect de contraintes temporelles et budgétaires.

## Ma mission chez Nomalys

Proposé une technique de gestion de projet et un logiciel adéquat à cette technique, ensuite étude et développement de la partie centrale de l'application NOMALYS qui consiste à remonter les informations nécessaires et pertinentes à l'utilisateur.

## Méthodologie de travail

Pour parvenir à proposer une technique de management qui correspond au besoin des projets dans cette entreprise et la mettre en place. Nous avons en premier lieu étudié les différentes méthodes existantes, ensuite nous avons mis en place une des techniques que nous avons jugées intéressantes à étudier.

Nous terminons notre travail avec la visualisation de notre technique à travers un projet qu'on a développé dans la société.

## Plan du mémoire

Le plan que nous adoptons dans ce manuscrit reflète les différentes évolutions de notre projet. Ce document comporte quatre chapitres. Après avoir présenté l'entreprise dans le premier chapitre, le deuxième chapitre étudié les différentes approches adoptées pour gérer un projet télécom. à partir de cette analyse, la seconde étape a été de mètre en place la politique de gestion dans l'entreprise.

Enfin une fois le choix de la technique est fait, nous nous sommes consacrés dans le dernier chapitre à exposer notre technique à travers un des projets que nous avons réalisé.

Nous résumons, en conclusion de ce manuscrit, les différentes contributions de ce projet.

Nous terminons notre travail avec la visualisation de notre technique à travers un projet qu'on a développé dans la société.

# Chapitre 1

# L'entreprise NOMALYS

## NOMALYS Smartphone

L'entreprise Nomalys est une startup de cinq personnes, deux associés et trois développeurs, elle a pour but de développer une application permettent d'accéder aux données d'entreprise depuis un Smartphone. À travers l'application NOMALYS l'utilisateur possède au bout de son pouce les informations qui feront la différence lors de rendez-vous et négociations.

## Motivation de création de la startup

### CRM

La gestion de la relation client est relativement récente. Le concept qu'il recouvre remonte au tout début des années 2000 et est le résultat des efforts commerciaux fournis par les sociétés de logiciels et de consulting pour promouvoir leurs solutions.

### Nomadisme

Au début du nomadisme les professionnels en commençait à adopter le CRM au PDA[1], avec l'apparition des Smartphones qui sont en même temps des téléphones et PDA, le basculement vers ces nouvelles gadgets de technologie était inéluctable.

### Intérêt pour les entreprises

Actuellement 51% des entreprises européennes estiment qu'elles ont raté des opportunités à cause du manque d'outils CRM mobile (relation client). Ce résultat, illustre l'importance croissante de ce type de solution au sein des entreprises (O Chicheportiche, 2009).

D'ailleurs, 49% des entreprises françaises interrogées ont déjà déployé ce type de solution. Pour autant, les déploiements concernent d'abord les grands comptes, les entreprises de moins de 500 salariés n'étant que 33% à déclarer avoir mis en place un projet CRM mobile.

### Un avantage concurrentiel

---

[1] Un **assistant numérique personnel** est un appareil numérique portable, souvent appelé par son sigle anglais **PDA** pour *Personal Digital Assistant.*

Ce que propose Nomalys est l'accès à aux données stratégiques tout le temps depuis n'importe où et que cet accès soit intuitif, rapide et sécurisé.

La recherche intuitive et suggérée permet de répondre immédiatement aux questions des clients sur le suivi de ses réclamations, le stock, sa commande, Lors d'âpres négociations, le commercial visualise en temps réel l'état de son compte client, ses délais de paiement.

# Le projet Nomalys

L'application NOMALYS permet un accès au CRM sans aucun logiciel à n'installer et aucune modification de système d'information actuel.

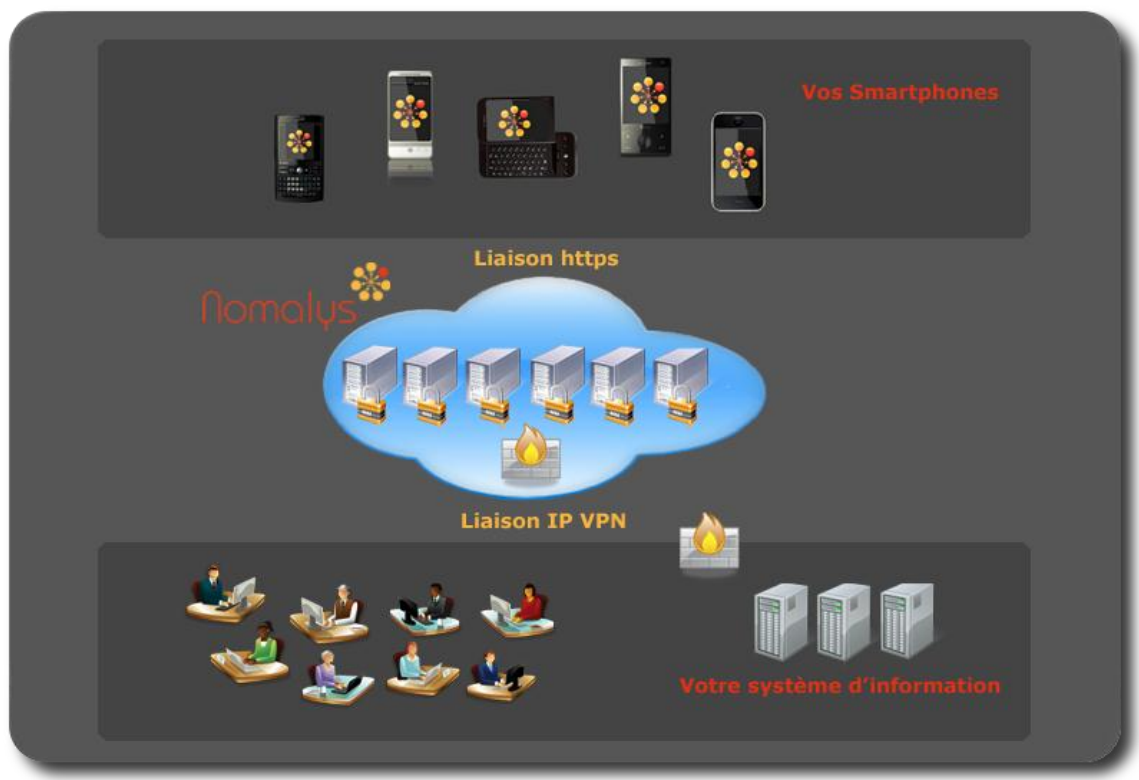

Figure 1 Déploiement de l'application NOMALYS [W1]

L'application Nomalys permet d'accéder au CRM et ERP avec une navigation intuitive :

- Rechercher simplement et rapide.
- Mettre en évidence les informations pertinentes.
- Répondre bien, vite et partout au client.

# Architecture de l'application Nomalys

Le schéma suivant représente les différentes couches de l'application Nomalys :

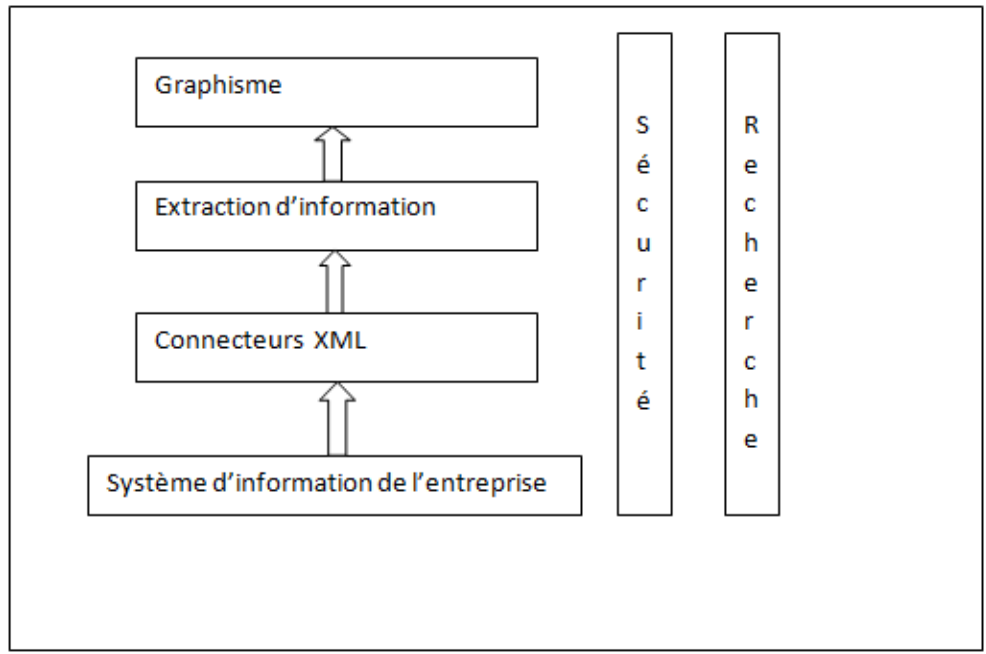

Figure 2  Architecture de l'application Nomalys

**Graphisme :** un désigne innovent inspirer du monde des jeux vidéo, 2 ans de recherche et développement et plusieurs brevets déposés, un travail avec des studios spécialisés dans l'innovation graphique et technologique autour des téléphones mobiles et plus particulièrement des Smartphones.

**Connecteurs XML** : Extraire les informations depuis le système d'information de l'entreprise l'application ce projet et affilier à un développeur de notre équipe.

**Extraction d'information :** Le bute est de remonter à l'utilisateur toutes les informations nécessaires et pertinentes, ce projet est sous ma responsabilité.

**Recherche d'information :** Moteur de recherche Google entreprise.

**La sécurité :** La connexion du système d'information de l'entreprise vers l'hébergeur français (Spécialiste Européen de l'infogérance d'applications IP à forte criticité) est totalement sécurisée via une connexion IP VPN. La connexion des serveurs Nomalys vers les Smartphones se fait via un protocole https.

# Conclusion

NOMALYS est une Startup innovation qui vient avec une nouvel approche du nomadisme avec l'approche désigne  venant du monde des jeux vidéo et de l'expérience de ces fondateurs dans le monde CRM, elle permet au utilisateur de manipulé son CRM agréablement et intuitivement, mais étant une startup qui vient de naitre elle ce devait de se munir d'une politique de gestion de projet puissante pour cadrer ses différents projets. Dans ce qui suis nous allons étudier les différents techniques de gestion de projet existâtes.

# Chapitre 2

# Les différents modèles de gestion de projet

## 1) Introduction

Dans ce chapitre, nous commençons par donner une brève historique de la gestion de projet. La section suivante est consacrée à l'étude de différentes techniques de management de projet informatique existante, nous détaillons ensuite les méthodes agiles qui sont une des tendances en ce moment en matière de gestion projet.

Nous terminons ce chapitre par une petite comparaissent des différents techniques.

## 2) Brève historique de la gestion de projet

Si la gestion de projet est généralement considérée comme une discipline moderne, il faut toutefois noter que la racine de ses principes fondamentaux provient de la fin du dix-neuvième siècle.

### 2.1) Les débuts : fin du dix-neuvième siècle

Si nous remontons à la dernière moitié du dix-neuvième siècle, à une époque où le monde des affaires commençait à devenir de plus en plus complexe, nous pouvons constater à quel point la gestion de projet a évolué depuis l'établissement des principes de gestion de base.

Ainsi, aux États-Unis, le premier grand projet réellement important qui a permis le penchant vert le problématique de la gestion de projet, fut la construction du chemin de fer transcontinental qui commença dans les années 1860. Brusquement, les chefs d'entreprise se sont trouvés face à la tâche impressionnante d'organiser le travail manuel de milliers de travailleurs(G. GREL,2003) .

### 2.2) Début du vingtième siècle

Avant la fin du siècle, Frederick Taylor (1856–1915) commença son analyse détaillée du travail. Il appliqua au travail un raisonnement scientifique en démontrant qu'il est possible d'analyser et d'améliorer le travail en le décomposant en parties élémentaires.

Auparavant, la seule méthode utilisée pour augmenter la productivité consistait à exiger des ouvriers un travail plus difficile et plus long. Taylor a instauré le concept d'un travail plus efficace à l'encontre d'un travail long et difficile *(F Kast,1960)*.

L'associé de Taylor, Henry Gantt (1861–1919) a étudié de manière approfondie l'ordre des opérations dans le travail, Il naitra de cela les Diagrammes de Gantt (affichage Diagramme de Gantt : affichage prédéfini qui affiche les tâches d'un projet sur le côté gauche de la vue et les barres graphiques correspondant aux durées de ces tâches sur le côté droit).

## 2.3) Milieu du vingtième siècle

Grâce à Taylor, Gantt et d'autres, la gestion de projet est devenue une fonction professionnelle à part entière qui nécessite des études et de la discipline.

Pendant la Deuxième guerre mondiale, des projets militaires complexes alliés à une offre de main-d'œuvre réduite en temps de guerre ont nécessité une réorganisation des structures. C'est à cette époque que furent créés les (organigramme des tâches : schéma montrant les relations entre les tâches de projet. Les tâches sont représentées par des cases, ou nœuds, tandis que les relations entre les tâches sont représentées par des lignes reliant les cases) (S.Aaron ,2004).

## 2.4) Aujourd'hui

De nos jours avec la diversification des projets, la gestion de projet a fait un vrai pas en avant et différents techniques de gestion projet sont apparut (Modèle en cascade, Modèle en V, Modèle en Spirale), toutes était développées pour palier à un vrai besoin de l'entreprise à gérer ses projets, dans ce qui suit nous allons les énumérer.

# 3) Modèles de développement

Dans ce qui suit nous allons présenter les principales techniques de gestion de projet qui ont existées dans le monde informatique.

## 3.1) Modèle en cascade

Hérité de l'industrie du BTP[2]. Ce modèle repose sur les hypothèses suivantes: On ne peut pas construire la toiture avant les fondations (R,Winston, 1970).

Le diagramme ci-dessous représentée le modèle en cascade :

---

[2] BTP : Bâtiment et des travaux publics

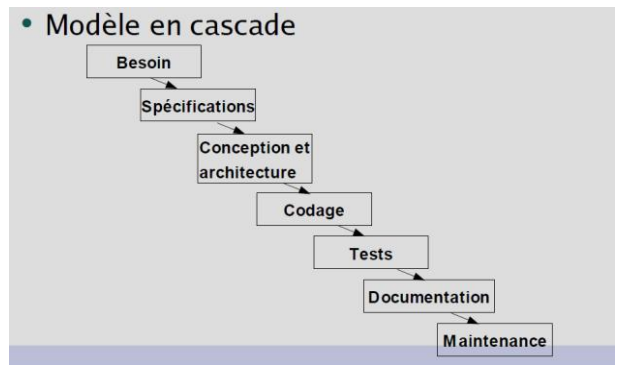

Figure 3 Modèle en cascade [W2]

Les phases traditionnelles de développement sont effectuées simplement les unes après les autres, avec un retour sur les précédentes, voire au tout début du cycle. Le processus de développement utilisant un cycle en cascade exécute des phases qui ont pour caractéristiques :

- de produire des livrables définis au préalable ;
- de se terminer à une date précise ;
- de ne se terminer que lorsque les livrables sont jugés satisfaisants lors d'une étape de validation-vérification.

## 3.2) Modèle en V

Le modèle du cycle en V est une amélioration du modèle en cascade qui permet en cas d'anomalie, de limiter un retour aux étapes précédentes (R.Winston, 1970).

Les phases de la partie montante, doivent renvoyer de l'information sur les phases en vis-à-vis lorsque des défauts sont détectés afin d'améliorer le logiciel, comme ces démontré dans le diagramme ci-dessous.

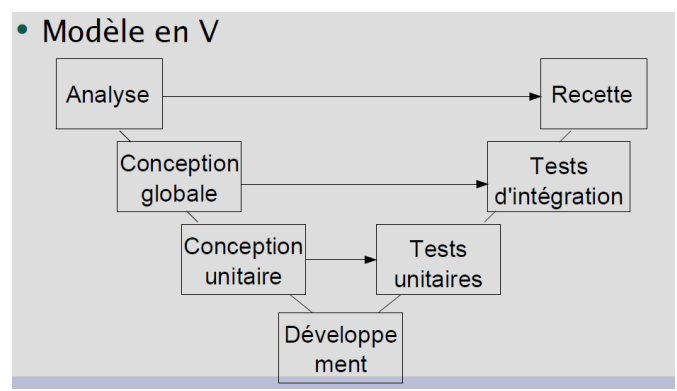

Figure 4 Modèle en V [W2]

De plus le cycle en V met en évidence la nécessité d'anticiper et de préparer dans les étapes descendantes les « attendus » des futures étapes montantes : ainsi les attendus des tests de validation sont définis lors des spécifications, les attendus des tests unitaires sont définis lors de la conception, etc.

Le cycle en V est devenu un standard de l'industrie du développement de logiciel et de la gestion de projet depuis les années 1980.

## 3.3) Modèle en Spirale

Le **modèle en spirale** (*spiral model*) est un modèle de cycle de développement logiciel qui reprend les différentes étapes du cycle en V. Par l'implémentation de versions successives, le cycle recommence en proposant un produit de plus en plus complet et dur. Le cycle en spirale met cependant plus l'accent sur la gestion des risques que le cycle en V (Boehm B, 1988).

Le diagramme ci-dessous représente le modèle en Spiral :

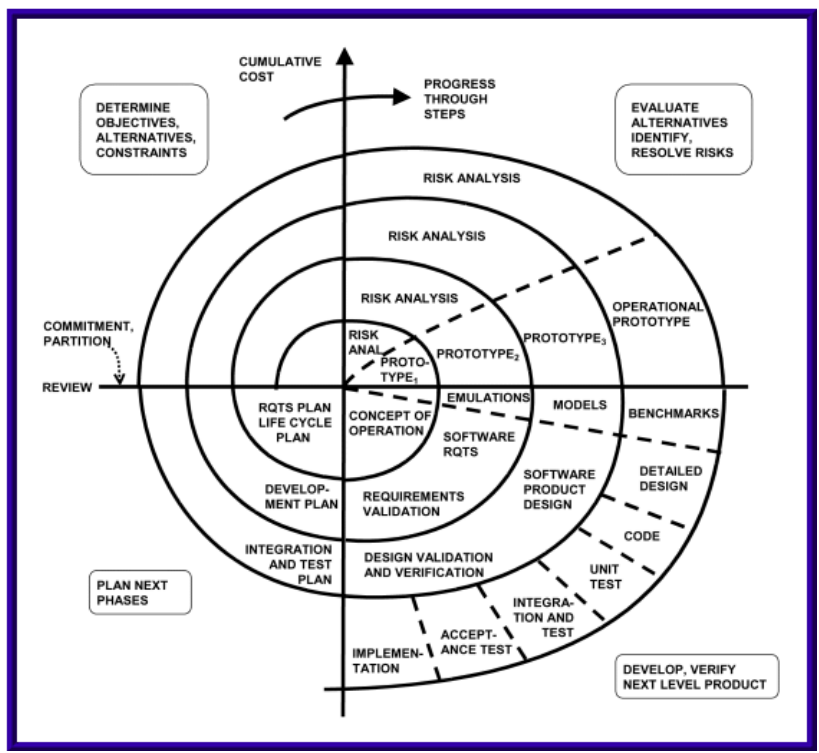

Figure 5 Modèle en Spirale [W3]

On distingue quatre phases dans le déroulement du cycle en spirale :

1. détermination des objectifs, des alternatives et des contraintes ;
2. analyse des risques, évaluation des alternatives ;
3. développement et vérification de la solution retenue ;
4. revue des résultats et vérification du cycle suivant.

## 3.4) Méthode agiles

Les méthodes de développement dites « **méthodes agiles** » visent à réduire le cycle de vie du logiciel (donc accélérer son développement) en développant une version minimale, puis en intégrant les fonctionn_è_alités par un processus itératif basé sur une écoute client et des tests tout au long du cycle de développement.

Les principes des méthodes agiles :

- individus et interactions plutôt que processus et outils
- développement logiciel plutôt que documentation exhaustive
- collaboration avec le client plutôt que négociation contractuelle
- ouverture au changement plutôt que suivi d'un plan rigide

## 4) Synthèse

Il n'y a pas de modèle idéal car tout dépend des circonstances. Le modèle en cascade ou en V est risqué pour les développements innovants car les spécifications et la conception risquent d'être inadéquate et souvent remis en cause. Le modèle incrémental est risqué car il ne donne pas beaucoup de visibilité sur le processus complet. Le modèle en spirale est un canevas plus général qui inclut l'évaluation des risques.

Souvent, un même projet peut mêler différentes approches, comme le prototypage pour les sous-systèmes à haut risque et la cascade pour les sous systèmes bien connus et à faible risque.
Les méthodes agiles quand a eux s'opposent aux méthodes conventionnelles (méthodes en cascade basées sur Merise, SADT et cycle en V) par leur forte itération et leur quasi-absence de formalisme.

## 5) Conclusion

Étant dans une petite startup de développement environnement évolutif ou le formalisme n'a pas un vrai impacte sur les projets et où la forte itération et nécessaire, nous avons décidé de développer nos applications on méthode agile. Le projet chapitre sera logiquement consacré à l'étude de ces méthodes.

# Chapitre 3

# La gestion de projet chez NOMALYS

## 1) Introduction

En a remarqué dans le chapitre précèdent que le choix de gérer le développement de l'application en méthode agile était indiscutable, notre prochaine étape sera d'étudier les différentes méthodes agiles et de trouver la méthode la plus approprier à notre développement, nous aurons ensuite à proposer un logiciel de gestion de projet qui se couplera avec notre technique de gestion de projet, permettant une bonne évolution de nos projets.

## 2) Choix de la méthode agile

### 2.1) RAD-Developpement rapide d'applications

La « **méthode de développement rapide d'applications** » (en anglais *Rapid Application Development*, notée *RAD*), consiste en un cycle de développement court basé sur 3 phases (Cadrage, Design et Construction) dans un délai idéal de 90 jours et de 120 jours au maximum.

### 2.2) DSDM

La méthode **DSDM** (*Dynamic Software Development Method*) a été mise au point en s'appuyant sur la méthode RAD afin de combler certaines de ses lacunes, notamment en offrant un canevas prenant en compte l'ensemble du cycle de développement.

Les principes fondateurs de la méthode DSDM sont les suivants :

- Une implication des utilisateurs
- Un développement itératif et incrémental
- Une fréquence de livraison élevée
- L'intégration des tests au sein de chaque étape
- L'acceptation des produits livrés dépend directement de la satisfaction des besoins

### 2.3) UP – Unified Process

La méthode du **Processus Unifié** (*UP* pour *Unified Process*) est un processus de développement itératif et incrémental, ce qui signifie que le projet est découpé en phases très courtes à l'issue de chacune desquelles une nouvelle version incrémentée est livrée.

Il s'agit d'une démarche s'appuyant sur la modélisation UML pour la description de l'architecture du logiciel (fonctionnelle, logicielle et physique) et la mise au point de cas d'utilisation permettant de décrire les besoins et exigences des utilisateurs.

## 2.4) RUP – Rational Unified Process

**RUP** (*Rational Unified Process*) est une méthode de développement par itérations promue par la société *Rational Software*, rachetée par IBM.
RUP propose une méthode spécifiant notamment la composition des équipes et le calendrier ainsi qu'un certain nombre de modèles de documents.

## 2.5) XP-extreme Programming

La méthode XP (pour *eXtreme Programming*) définit un certain nombre de bonnes pratiques permettant de développer un logiciel dans des conditions optimales en plaçant le client au cœur du processus de développement, en relation étroite avec le client.
L'eXtreme Programming est notamment basé sur les concepts suivants :

- Les équipes de développement travaille directement avec le client sur des cycles très courts d'une à deux semaines maximum.
- Les livraisons de versions du logiciel interviennent très tôt et à une fréquence élevée pour maximiser l'impact des retours utilisateurs.
- L'équipe de développement travaille en collaboration totale sur la base de binômes.
- Le code est testé et nettoyé tout au long du processus de développement.
- Des indicateurs permettent de mesure l'avancement du projet afin de permettre de mettre à jour le plan de développement.

## 2.6) Crystal clear

C'est une méthode de gestion de projet. Sa méthodologie est très fortement adaptable aux spécificités de chaque projet. Plusieurs principes doivent être partagés par l'ensemble de l'équipe :

- La communication est omniprésente pour réussir le « jeu coopératif » que représente un projet comme le fait remarquer le créateur de cette méthode : Alistair Cockburn.
- Le nombre de membres d'une équipe est limité à six personnes afin que l'équipe soit solidaire.
- Tous les membres de l'équipe doivent travailler dans une même pièce afin de faciliter la communication par proximité.
- Les schémas de modélisation doivent être réalisés en groupe et sur tableau blanc car cela améliore la communication et la collaboration.
- La collaboration avec le client est elle aussi très importante, notamment grâce à de nombreuses conversations entre utilisateurs et développeurs.
- Livrer des parties exécutables de l'application le plus fréquemment possible afin que le client se rende compte du travail en cours et propose des changements.

## 2.7) Étude comparatif

Pour pouvoir choisir notre technique de développement, nous avons cherché à trouver le lien qui avait entre les méthodes agiles et l'environnement du projet qui veut dire le nombre de personne impliquait dans le projet est la taille du projet.

Ce schéma représente l'application des méthodes agiles par rapport à la taille des projets.

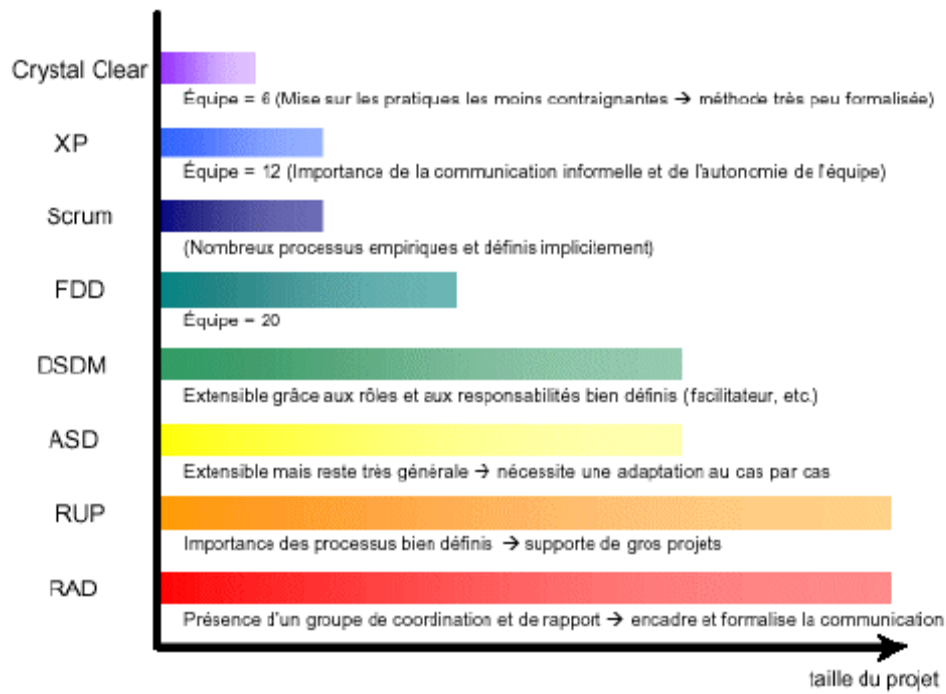

**Figure 6 Méthodes agiles [W4]**

Nous remarquons que par rapport à ce schéma nous nous situent dans le bandeau violé avec un nombre de personne très réduit dans le projet pas plus de 6 personnes, cela correspond à l'application de la méthode agile « Crystal clear ».

Dans ce qui suit nous allons montrer comment on a instauré la méthode crystal clear dans le management de nos projets.

## 2.8) Descriptif de notre choix on technique de gestion de projet

Crystal clear est une méthode de gestion de projet. Sa méthodologie est très fortement adaptable aux spécificités de chaque projet. La communication est omniprésente pour réussir le « jeu coopératif » que représente un projet.

Plusieurs principes doivent être partagés par l'ensemble de l'équipe, nous montrons ci-dessous comment nous les respectons :

### 2.8.1) Condition 1

- Le nombre de membres d'une équipe est limité à six personnes afin que l'équipe soit solidaire.
- Nous sommes une équipe de quatre personnes.

### 2.8.2) Condition 2

- Tous les membres de l'équipe doivent travailler dans une même pièce afin de faciliter la communication par proximité.
- Nous sommes dans une pièce de l'incubateur de l'INT.

### 2.8.3) Condition 3

- Les schémas de modélisation doivent être réalisés en groupe et sur tableau blanc car cela améliore la communication et la collaboration.
- Les réunions ce passe dans une salle de réunion ou chaque membre de l'équipe donne son avis sur le chemin que doit suivre le projet.

### 2.8.4) Condition 4

- La collaboration avec le client est elle aussi très importante, notamment grâce à de nombreuses conversations entre utilisateurs et développeurs.
- Dans notre entreprise on peut considérer que le client c'est le manager, puisque c'est lui qui fait les spécifications de l'application.

### 2.8.5) Condition 5

- Livrer des parties exécutables de l'application le plus fréquemment possible afin que le client se rende compte du travail en cours et propose des changements.
- Avec Redmine le client peut avoir une session lui permettent de voire l'avancement de toutes les demandes.

### 2.8.6) Condition 6

Crystal reste très souple tant au niveau des procédures à suivre que des normes à utiliser (comme par exemple les normes de codage). Cette méthode possède une procédure découpée en différentes étapes :

6.1)

- La spécialisation consiste à observer les utilisateurs dans leur travail pour mieux connaître leurs besoins et leur environnement. Ensuite, les différents cas d'utilisation sont classés par ordre de priorité en collaboration avec les utilisateurs, ce qui permet de savoir quelles fonctionnalités ont le plus de valeur et doivent être développées en premier.
- Dans notre ca c'est le manager qui défini les spécifications.

6.2)

- Le planning consiste à prévoir vers quelles dates les itérations vont se suivre, il est recommandé de définir des itérations d'une longueur de 2 à 3 mois, chacune produisant un produit à livrer fonctionnel.
- A NOMALYS c'est le manager qui définie des sprinte de (un à deux) semaines avec forcément un produit après chaque sprinte.

## 2.9) Conclusion

Crystal clear présente tout les avantages des méthodes agiles : flexibilité par rapport au changement, rapidité, livraisons fréquentes, etc. Elle convient tout à fait pour une petite structure (taille inférieure à 6 personnes) comme la notre.

# 3) Choix de l'outil de gestion de projet

Après avoir choisi une technique de management de nos projets nous nous devient de nous prémunir d'un outil ayant pour objectif de faciliter le travail de gestion de projet.

## 3.1) Définition

Le travail des logiciels de gestion de projet est généralement d'automatiser des tâches de sauvegarde et/ou de la gestion du temps. Par exemple d'enregistrent différents états d'un projet et gardent une trace de la date de modification.

## 3.2) L'offre du marché

Pour faire notre recherche nous nous sommes fait aider par les sites des éditeurs logiciels et aussi par des blogues qui nous on permit d'avoir de vrai critique sur les différents produits.

Partent de ça nous avons distingué deux types de produit :

### 3.2.1) Solutions propriétaires

De nombreux outils existent sur le marché, en tête du marché l'application Microsoft Project qui permet de planifier les projets et les ressources, et d'assurer le suivi des projets pendant leur réalisation. MS Project permet au chef de projet d'assurer une gestion de projet professionnelle, conforme à l'état de l'art, et ainsi de garantir le respect des délais et du budget.

Les grandes fonctions de la version EPM (Enterprise Project Management):

- Planification et gestion de projets
- Gestion de portefeuilles de projets
- Gestion des ressources
- Travail collaboratif

### 3.2.2) Solutions open source

Les acteurs Open Source commence à apparaître et à se faire une place sur le marché. S'ils sont encore peu nombreux, certains proposent des fonctionnalités aussi complètes que les solutions propriétaires

Dans ce tableau nous avons fait un récapitulatif des différents produits qui nous intéressent :

| Les principales solutions de gestion de projets Open Source | | | |
|---|---|---|---|
| **Solution** | **Version et technologie** | **Type - Origine** | **Principales fonctionnalités** |
| **Dotproject** | Linux, Windows ● Apache ● MySQL, PHP | Solution de gestion de projets avec des fonctionnalités Groupware | Gestion de projets, des tâches, des dépendances ● Gantt ● calendrier ● forum ● Gestion des agendas ● annuaires ● modules entreprises, clients |
| **Gantt project** | Version 1.10 (08/2004) ● Dvp Java ● GNU /Linux, Windows, MacOsX. | Planification de base ● Initialisé par un français. | Gestion des tâches, ressources et dépendances ● Exportation HTML ou en images PNG ● Simple d'utilisation, qualité du graphisme. |
| **Imendio Planner** | Version 0.12.1 (08/2004) ● Linux, BSD et UNIX | Planification et suivi de projets pour Gnome 2 ● Anciennement MrProject de Codefactory | Gantt ● définition des tâches et sous tâches ● des ressources ● des dépendances entre tâches ● chemin critique ● vue d'ensemble sur l'utilisation des ressources ● exportation HTML ● disponible en Français |
| **Iteamwork** | - | Planification de base - gestion de projets en ligne ● Site américain | Gestion de projets, des tâches ● gestion des ressources ● dépendances ● calendrier ● progression ● notification par email |
| **Open Workbench** | Version 1.0 ● Java and C++ ● MS Windows | Vise des besoins de planification élevés ● Editeur californien Niku | Gestion de projets, sous-projets, ressources, tâches ● gestion des dépendances ● gestion des calendriers ● Gantt, Pert ● automatisation ● cycle de vie, estimations, progression ● WBS ● chemin critique ● disponible en français. |
| **Redmine** | Linux ● Apache ● PHP et MySQL. | Solution Groupware - gestion de projets en ligne ● Editeur français | Planification et gestion des tâches, des ressources ● Cycle de vie du projet ● notifications email ● graphiques et statistiques ● forum ● calendrier, partage des documents ● accès client |

Tableau 1 Solution de gestion de projet open source

## 3.3) Critères de sélection pour le logiciel

Il est difficile de comparer ces services tant les fonctionnalités peuvent varier. On retrouve généralement une liste de tâches et un calendrier, mais certains vont intégrer un outil pour suivre les bugs.

De la même manière, leur mode de fonctionnement (hébergés par l'éditeur ou à héberger soi-même sur ses propres serveurs) sont très disparates. Les technologies employées peuvent également rentrer en ligne de compte dans le choix.

De notre part, nous étions à la recherche d'un logiciel qui remplisse les critères de sélection suivant :

**Logiciel gratuit :** Étant une Start up, nous étions à la recherche d'un logiciel gratuit puisque l'entreprise n'a pas de grosse ressource financier pour se permettre un des logiciels payants.

**Open source :** c'est-à-dire la possibilité d'accès et de modification du code source.

**Application web :** nous préférons une application web permettent un accédé distant, plus facile pour des manager nomade et aussi pour le client qui pourra avoir une possibilité de suivi des projets à distance.

**Langue française :** de préférence puisque nous sommes en France et toute notre équipe est francophone.

## 3.3) Choix Redmine

Du fait des différents critères de sélection citait au par avons, on peut dire que Redmine répond à nos critères (mode web, open source, multi langue dans le français et aussi gratuit), un autre critère qui a joué aussi en sa faveur c'est le fait que les autres startups qui sont incubés avec nous, nous on fait de bons retours sur le produit.

### 3.3.1) Principales fonctionnalités

**Gestion multi-projets:** Avec Redmine l'utilisateur à la possibilité de géré une infinité de projet.

**Gestion fine des droits utilisateurs :** Ces droits sont décrits dans l'application par des rôles ou par profil utilisateur que l'administrateur affiliera a u utilisateur.

Planification et gestion des tâches, des ressources

**Notifications email :** L'application peut être configurée pour envoyer un email au manager pour l'informer de l'avancement des projets.

**Graphiques et statistiques :** l'application permet de faire des rapports statistiques sur divers critères.

**Forum :** Avec le forum les développeurs peuvent se partager des informations pertinentes en instantané sur un Bugue ou un problème donné.

**Partage des documents :** Redmine permet un facile partage des données que les utilisateurs peuvent exploiter dans leurs projets.

**Accès client :** l'application Redmine peut être configurée pour permettre au client de suivre le développement du projet.

## 3.3.2) Installation de Redmine

Nous avons installé notre application sur notre serveur distant OVH[3] , sous un système d'exploitation Windows, nous y accédons sur le port 3000 du même serveur, pour cela nous avons eu à installer des logiciels pré requit par exemple :

**Mysql 4:** pour pouvoir gérer notre base de données.

**PHP 5 :** pour pouvoir développer d'autre fonctionnalité si nous avons besoin.

**Phpmyadmin :** pour pouvoir faire des sauvegardes.

## 3.2.1) Gestion du temps avec Redmine

Avec Redmine, chaque utilisateur doit à chaque fin de journée détailler le temps passée dans chaque projet.

Les développeurs doivent insérer le nom du projet, leur activité et le nombre d'heure passé dedans. La figue ci-dessous représente l'interface qu'à l'utilisateur pour remplir ces informations.

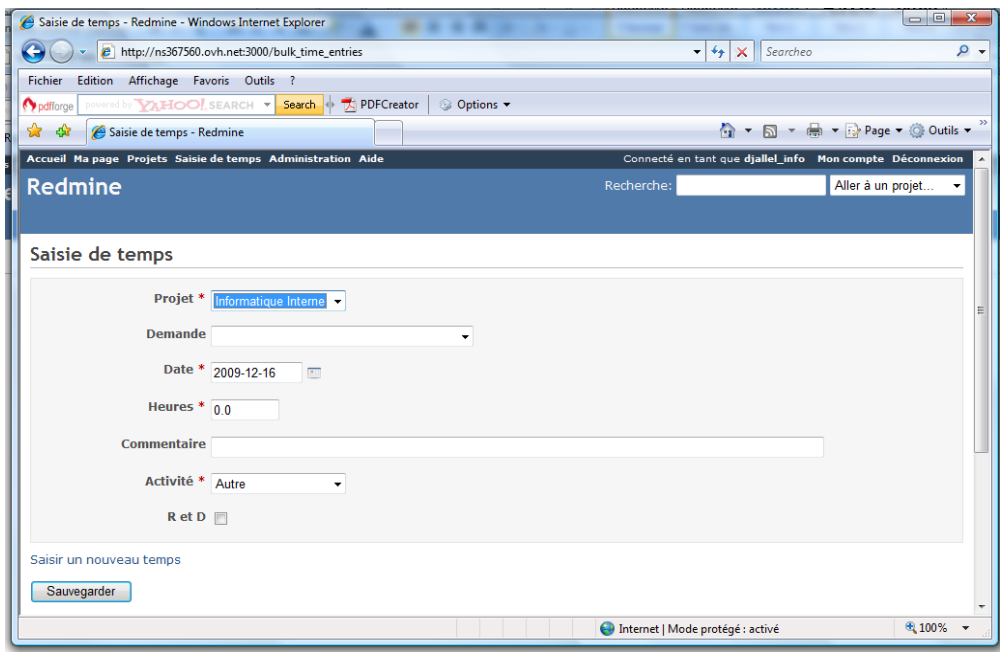

Figure 7 Saisie de temps sur Redmine

---

[3] OVH est un hébergeur de sites internet français, orienté grand public. Il propose des serveurs dédiés, des serveurs privés, de l'hébergement mutualisé, l'enregistrement des noms de domaine (sous l'intitulé Registrar), ainsi que de la téléphonie sur IP.

## 1-Diagramme de Gantt

Avec le diagramme de Gantt Le manager peut à tout moment observer l'avancement du projet.

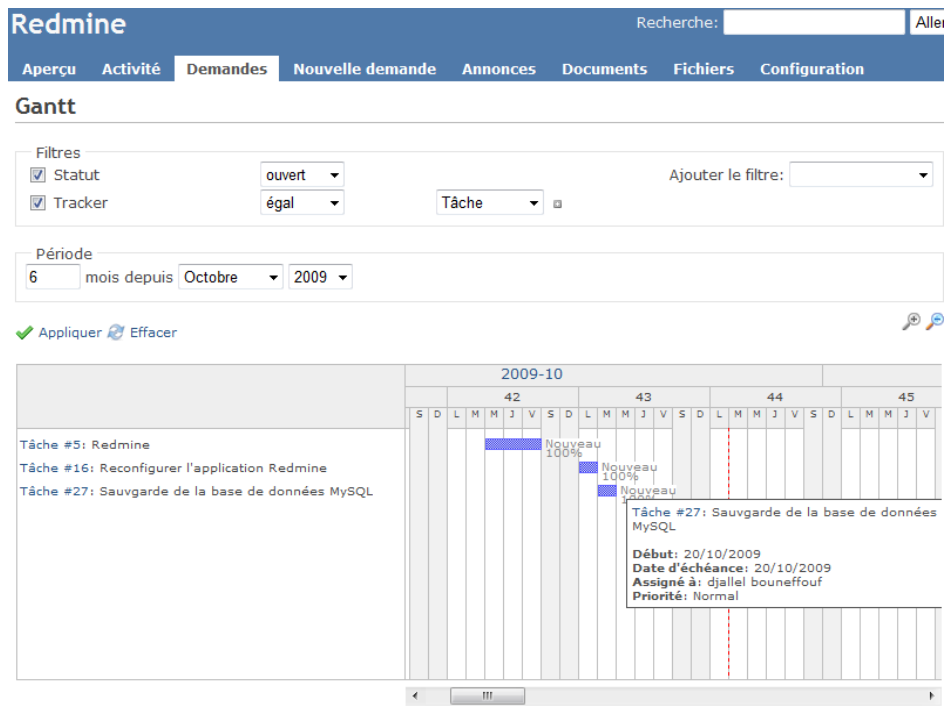

Figure 8 Diagramme de gantt

Comme c'est illustré dans la figue ci-dessus, l'utilisateur à une interface lui permettent de suivre toutes les tâches dans l'intervalle de temps qu'il aura choisi, aussi la possibilité de voir la date de début, la date d'échéance et la priorité de la tâche.

## 3.2.1) Gestion des connaissances avec Redmine

### 1-Définition

La gestion des connaissances (en anglais *Knowledge Management*) est l'ensemble des techniques permettant de mémoriser et de partager des connaissances entre les membres des organisations, en particulier les savoirs créés par l'entreprise elle-même en vue d'atteindre l'objectif fixé.

### 2-Gestion de la connaissance chez Nomalys

Avec la masse d'information qui commence à s'accumuler au tour de nos projets nous devions de mettre en place un système de gestion de nos connaissances internet que se sois au niveau de nos techniques de développement ou dans les rapports que nous gênerons.

**Library de programme :** à chaque foi qu'on développe une application nous la mettons dans une librairie dans le serveur local de l'entreprise pour permettre aux autres développeurs de l'exploiter.

En plus du serveur local partagé qui contient tous les projets et les documentations de nos projets, Toutes la documentation du code que les développeurs ont rédigé sera stocké dans Redmine. Ci-dessus

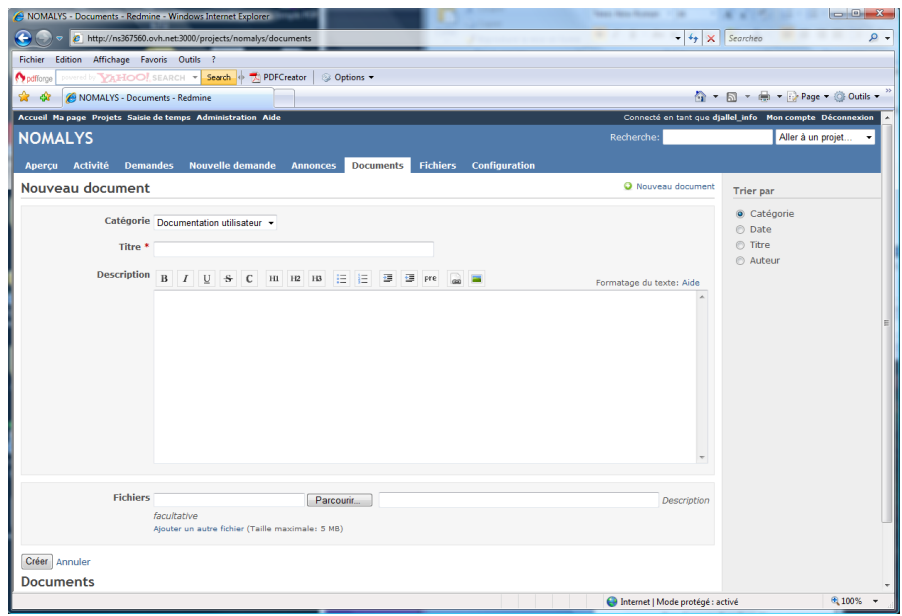

Figure 9 Insertion des documents sous Redmine

## 3.2.1) Méthode Agile avec Redmine

Nous pourrons dire que le développement agile est architecturé sur trois étapes comme c'est montré dans le diagramme ci-dessous :

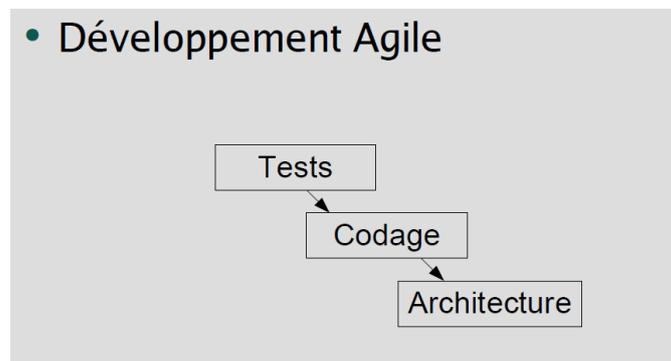

Figure 10 Méthode agile [W2]

Dans ce qui suit nous décrivons comment nous avons utilisé Redmine pour gérer nos projet on méthode agile.

**1-Création des tests selon les spécifications**

**Le manager :** crée les spécifications de l'application et la transmet aux développeurs par une demande depuis l'application Redmine,

**Le développeur :** écrie les testes auxquels l'application doit répondre.

## 2-Codage en conformité aux tests

Chaque développeur doit faire une uploade de son application. Dans la partie fichier de Redmine comme c'est montré dans la figure ci-dessus.

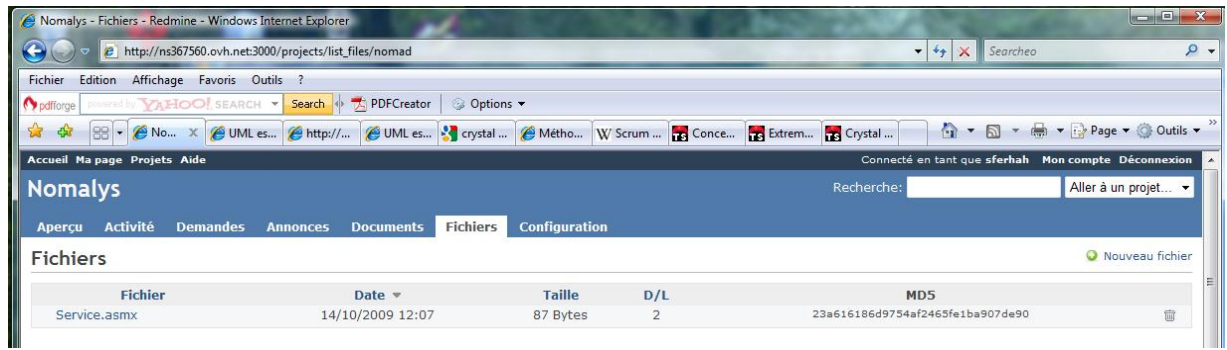

Figure 11 Insérer des fichiers sous Redmine

Cette opération permettra d'abord un archivage du code mais, aussi la possibilité au manager de vérification le bon fonctionnement de l'application.

## 3-Émergence de l'architecture

Chaque développeur uploade dans Redmine son rapport de la semaine, qui doit contenir la modélisation UML de l'application qu'il a développée.

Les modélisations UML qui doivent être uploadé son les suivants :

- **Diagramme d'objet:** Généré de puis le code, donne une vu global du code.
- **Diagramme de séquence:** Permet une vue temporelle du défilement du code.
- **Diagramme de cas d'utilisation:** Permet de voir les interactions avec l'utilisateur.

La figure suivante montre l'interface Redmine où l'utilisateur uploade ses rapports.

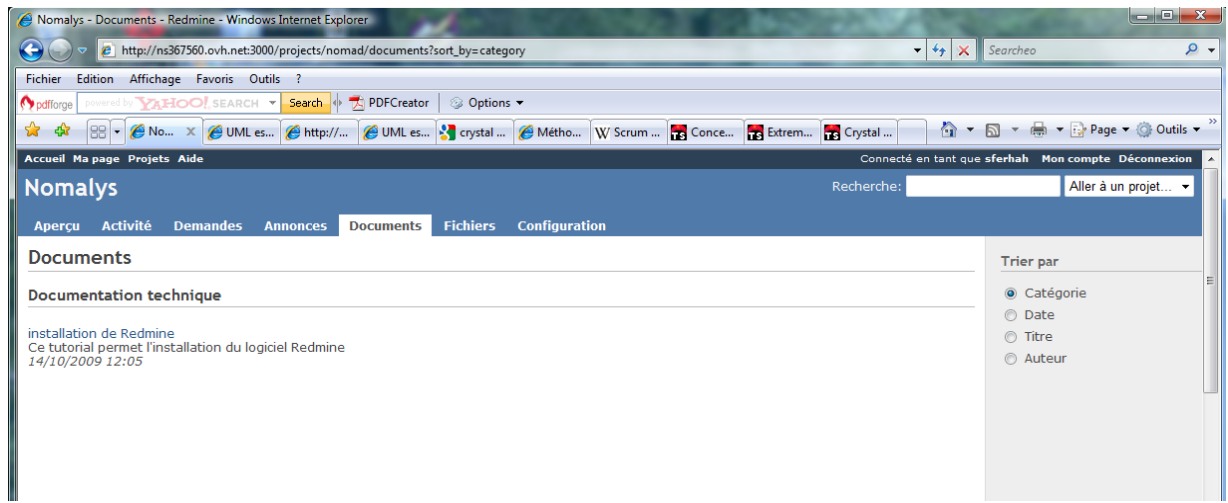

**Figure 12 Documentation technique**

# 5.6) Conclusion

Dans ce chapitre nous avons exploré les différentes techniques agiles (sacrum, Crystal clear, XP-extrême Programming, Unified Process) qui nous a permis de choisir la méthode la plus approprier à appliquer dans nos projets, nous avons ensuite décrit comment nous avons essayé de respecter principe de la technique dans notre équipe.

La deuxième partie du chapitre était consacré à l'étude du marché des logiciels de gestion de projet qui nous a orienté vert un bon logiciel de gestion projet qu'est Redmine.

Enfin nous nous sommes focalisés à décrire comment Redmine nous a permis de gérer nos projets. Nous allons décrire dans le prochain chapitre un des projets que nous avons eu à gérer,

# Chapitre 4

# Gestion du projet filtrage

## 1) Introduction

Après avoir établi une politique de gestion de projet, on peut commencer à développer l'application NOMALYS dans de bonne condition.

En premier temps on a élaboré un l'état de l'art des techniques existantes en matière de filtrage et on a proposé une technique de filtrage, la suit logique était de développer plusieurs versions de l'application de filtrage.

Dans ce qui suit nous allons présenter les différentes étapes du projet qu'on a eu à développer.

## 2) État de l'art

Dans cette étape nous avons eu 15 jours pour pouvoir étudier le filtrage d'information et les différentes techniques existantes et proposer une technique qui va permettre de répondre à nos besoins en termes d'extraction d'informations pertinentes.

### 2.2.1) Cahier des charges

Ce projet s'insère dans le cadre du filtrage de l'information appliquée au système d'information mobile. Il a pour but de développer un système de filtrage d'information permettant de remonter les informations les plus pertinentes à des utilisateurs nomades.

Cette étude permettra de dégager des techniques de filtrage d'information approprié au monde des systèmes d'information mobiles.

### 2.2.1) Filtrage

Au lieu de laisser l'utilisateur dépenser son temps à chercher l'information dont il a besoin, la tendance actuelle est de concevoir des mécanismes qui permettent de lui faciliter la tâche en lui faisant parvenir continuellement l'information qui l'intéresse.

### 2.2.2) Méthodes de Filtrage

Vu que le processus de filtrage est relativement analogue à la recherche d'information, ce sont les mêmes techniques de recherche d'information qui ont été adoptées à ce filtrage et c'est seulement l'approche ou la vision qui diffère.

**Filtrage Full-text** : C'est une méthode directe qui consiste, en se basant sur le parcours du texte, à sélectionner tous les documents contenant une certaine chaîne de caractères (mots clés, expression booléenne,...).

**Filtrage basé sur l'indexation** : Chaque document est représenté par une liste de mots clés décrivant le contenu du document. Les mots clés sont stockés dans un fichier indexe et pour chaque mot clé est établie une liste de pointeurs désignant les documents qui lui sont relatifs.

**Filtrage Booléen** : L'utilisateur exprime ses profiles par des mots qui doivent exister ou ne doivent pas exister dans le document à recevoir. Le modèle n'autorise que la conjonction et la négation des mots. Cependant, l'utilisateur peut simuler la disjonction en plusieurs profils.

## 2.2.3) Proposition d'une première architecture de l'application filtrage

La première étape consiste à bien comprendre le système à étudier pour mieux délimiter le système à étudier. La méthode générale qu'on a adopté consiste à :

Retrouver les acteurs qui interagissent avec le système.

Rechercher les fonctionnalités du système par l'utilisation des « cas d'utilisation »

Le schéma ci-dessus présente l'architecture de notre système de filtrage :

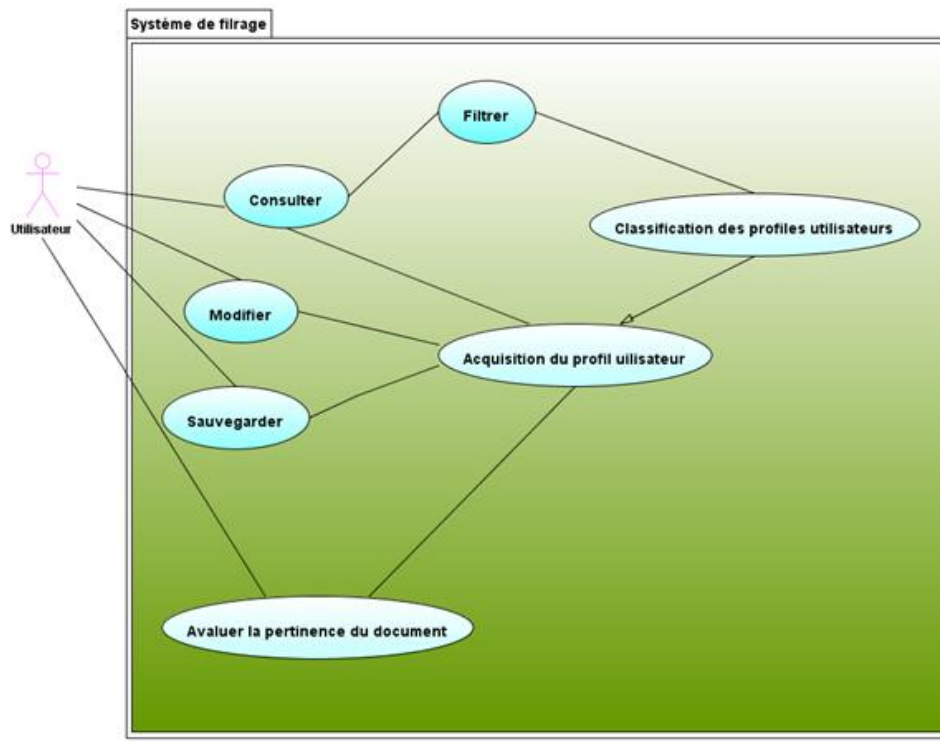

Figure 13 Système de filtrage

**Les actions de l'utilisateur :** consulté, modifié, sauvegardé.

**Fonction de l'application filtrage**

- Évaluer la pertinence du document
- Acquisition du profil utilisateur
- Classification des profils utilisateur

# 3) Architecture

Après une réunion stratégique nous avons changé les objectifs et nous nous sommes rediriger vers une nouvelle architecture de la partie filtrage et nous avions 15 jours pour développer la première version de l'application filtrage.

## 2.2) Travail réalisé

Dans cette partie nous allons à proposer une architecture globale de l'application Nomalys intégrant notre système.

Si dessous est représentée l'architecture de l'application Nomalys.

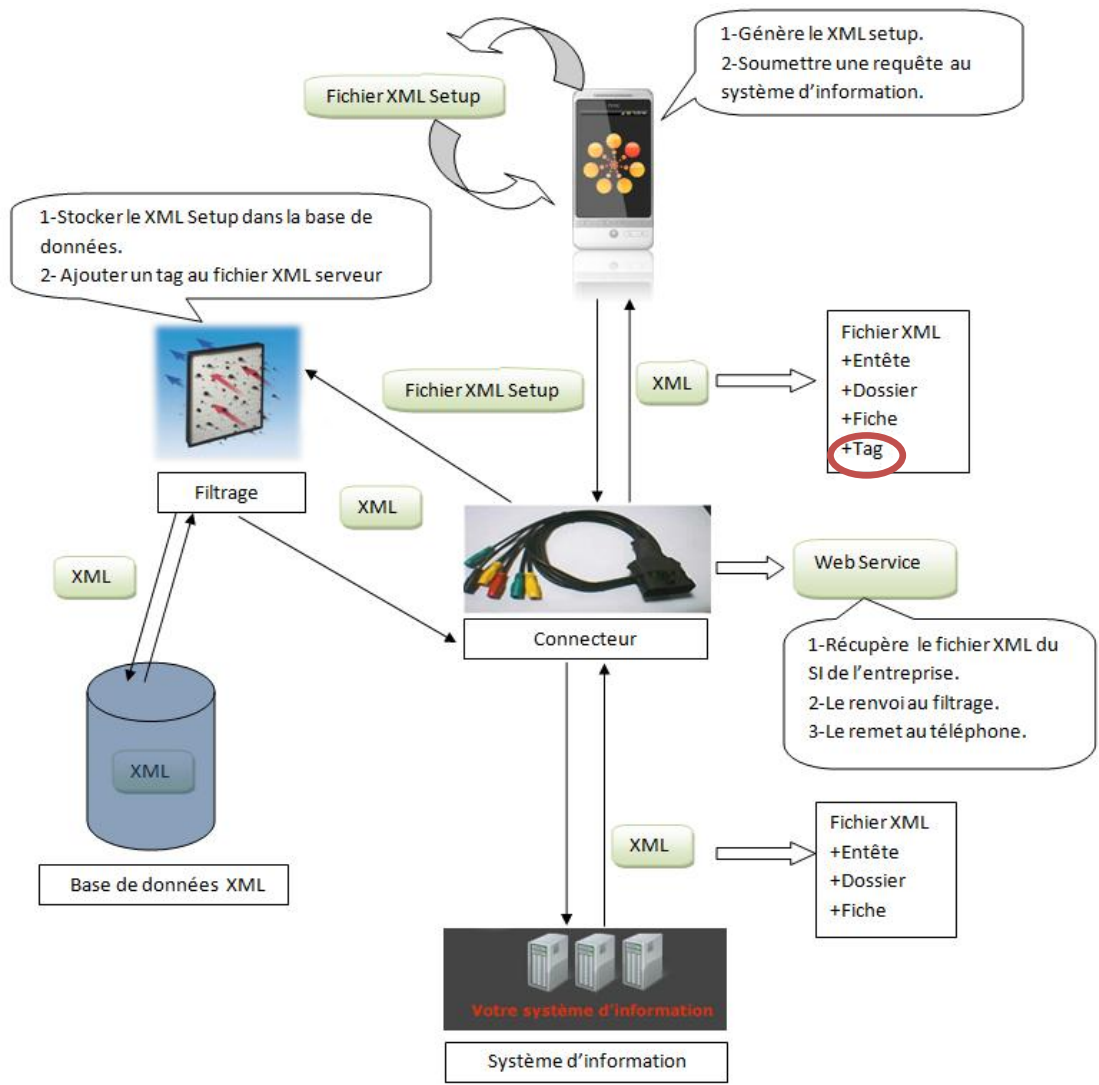

Figure 14 Architecture globale de Nomalys

## 2.2.1) Architecture

L'application Nomalys est architecturée comme suit :

Deux composent principaux :

### 1-Connecteurs

Le connecteur permet d'extraire les informations depuis le système d'information de l'entreprise l'application et les remonter vert le smarthphone.

### 2-Filtre

Permet de remonter à l'utilisateur toutes les informations nécessaires et pertinentes.

Pour le faire on utilise deux composent :

**Le Setup** : Le fichier setup est créé dans le téléphone, il contient les champs visibles et leurs positions choisi par l'utilisateur, ainsi que les champs non visibles, ce fichier va être stocké dans une base de donné et il sera renvoyé à l'utilisateur laure de ça prochaine connexion à l'application.

**Base de données XML / USER :** La base de données XML doit contenir le profil de tous les utilisateurs.

# 4) Développement

Avec la méthode agile nous sommes partis sur une base de proposer après chaque fin de sprint une application qui fonctionne même si celle-ci elle contient peu de fonctionnalité ou pas les fonctionnalités attendues par le client.

## 4.1) Base de données XML 0.1

Cette étape à durée 15 jours où nous avons eu à développer deux web services qui communiquent entre eux, un qui contient le filtre et l'autre qui contient le connecteur.

Le diagramme ci-dessus représente bien la communication entre le deux web service.

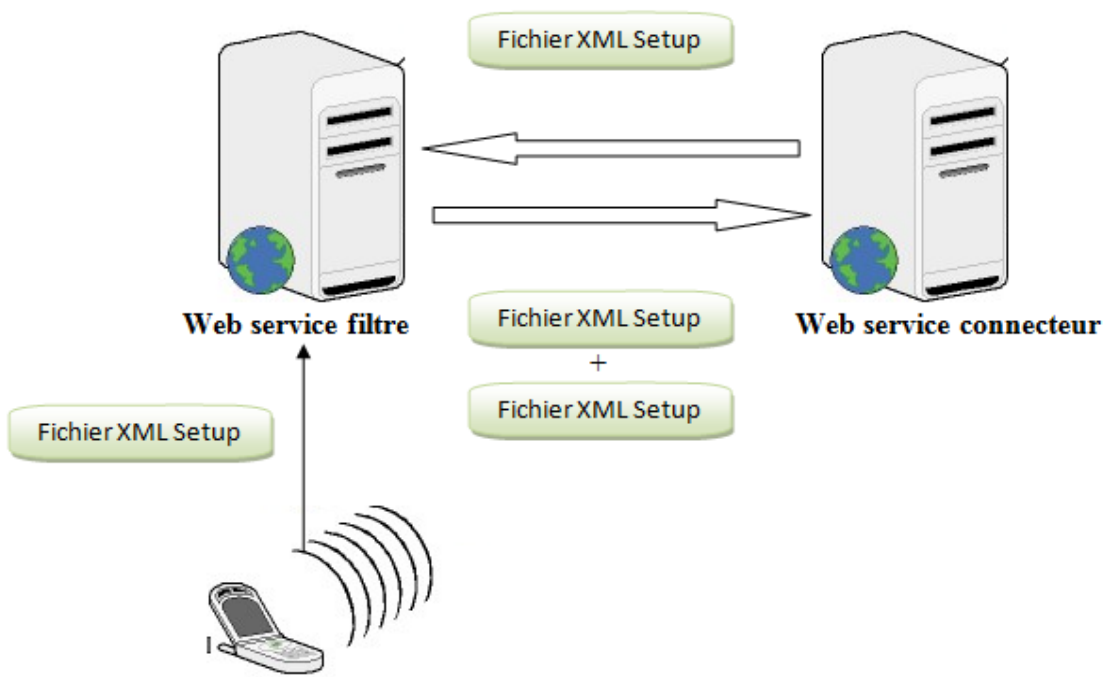

Figure 15 Communication entre webservices

### 4.1.1) Web service connecteur

Contient les deux applications

**Envoisetup :** Envoie le setup reçu par le téléphone vers le « web service filtre ».

**EnvoiFichierXML :** Envoie le fichier XML vert le « web service filtre » et il le récupère tagger avec un profil utilisateur.

### 4.1.2) Web service filtre

Contient les deux applications

**Insertsetup :** Insert le setup reçu par le « Web service connecteur » dans la base de données XML.

**Inserttag :** Insert le tag de l'utilisateur dans le fichier reçu par le web service numéro 1 et le lui renvoi.

Le « web service filtre » va comparer le user du fichier XML avec les user stocké dans ca base de données avant de lui insérer le tag.

## 4.2) Base de données XML 0.2

Dans cette étape nous remarquant que le cahier des charges à encore changé, nous avons ajouté une base de donné historique qui va contenir l'historique de déplacement des utilisateurs dans le téléphone.

Vu qu'on est on méthode agile le développeur accepte toujours l'évolution du cahier de charge dans 15 j de plus pour développé cette nouvelle application.

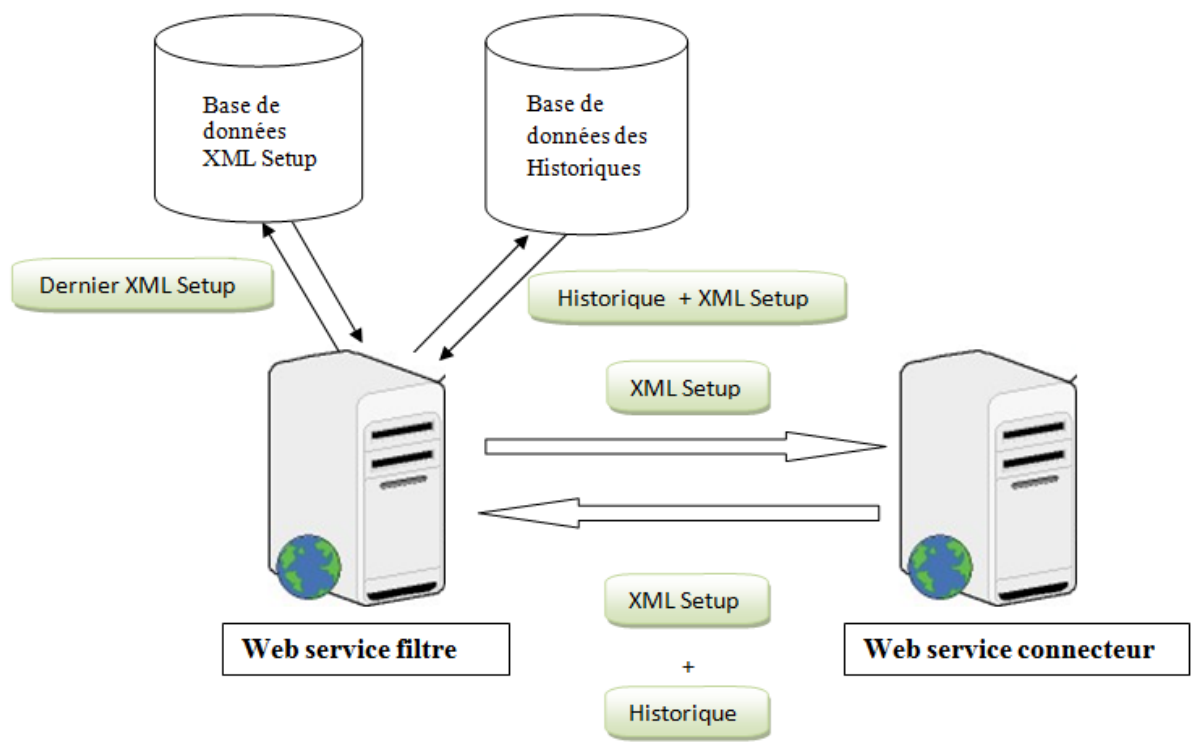

Figure 16 Webservice filtrage connecté à deux base de données

Comme c'est décrie dans le schéma ci-dessus le web service filtre contient deux base de donné :

**Historique :** déplacement des utilisateurs dans le Smartphone.

**Base de donné xml :** contient les informations que l'utilisateur veut afficher dans son téléphone.

## 4.3) Basculé vert une base de données XML vert une base de données SQL serveur

Dans cette étape j'ai un Sprinte deux 15 jours pour pouvoir étudier le filtrage d'information et les différentes techniques existantes et proposer une technique qui va permettre de répondre à nos besoins en termes d'extraction d'informations pertinentes.

## 4.4) Basculé un seul web service et des DLL

Après avoir fait marcher l'application en 4 web service, nous avions l'aspect temporel qui nous fessait défaut, puisque avec l'utilisation des web service nous étions à un temps de réponse au tour de 5 secondes. Plus de l'aspect temporelle l'utilisation des web service constitue une faille de sécurité pour notre application, du fait des précédent remarque nous avons choisi de regroupé les 4 web service en un seul web service qui applé des DLL[4].

## 4.5) Teste de l'application

Après avoir développé l'application nous passons logiquement à l'étape de test, cette étape permettra d'évaluer notre application en termes de temps de réponse et aussi en capacité de surcharge.

- Dans cette étape nous avons développé une application qui se connecte à notre web service et le Bombard de requête, cela pour savoir le nombre de requête simultané qui sont supportable par l'utilisateur.

- Aussi l'application va envoyer des requêtées à intervalle régulier, pour savoir l'intervalle supportable par notre web service.

# 5) Synthèse

Nous remarquons qu'au file du projet nos choix en matière de technique et d'architecture on varié plusieurs fois et cela nous permet de dire que dans notre projet la méthode agile nous a était vraiment bénéfique dans le sens ou nous n'avions pas un cahier des charges prédéfini.

---

[4] *DLL* signifie *Dynamic Link Library*, ou en français Bibliothèque de liens dynamiques, dans le cadre du Système d'exploitation Windows, et compatibles (wine, ReactOS). Traditionnellement, le nom de ces fichiers se termine par l'extension « .dll ». Une DLL peut contenir du code ou des ressources qui sont alors rendus disponibles à d'autres applications.

# Conclusion

Nous avons présenté, tout au long de ce manuscrit, notre démarche pour l'étude, la proposition et l'évaluation d'une technique de gestion de projet.

Pour cela nous avons d'abord étudié différente technique de management de projets informatiques existants. Cette étape nous a permis de constater qu'il n'y a pas de modèle idéal en termes de gestion de projet, car tout dépend des circonstances.

Vu que nous étions dans environnement évolutif ou le formalisme n'a pas un vrai impacte sur les projets et où la forte itération et nécessaire, nous avons décidé de développer nos applications on méthode agile.

Nous avons ensuite étudié les différentes méthodes agiles. Cela nous a permis de trouver la méthode la plus approprié à notre développement.

Nous avons ensuite étudié le marché des logiciels de projet et proposer un logiciel de gestion de projet, qui s'est bien couplé avec notre technique de gestion de projet, permettant une bonne évolution de nos projets.

Après avoir établi une politique de gestion de projet, nous avons commencé le développement de l'application NOMALYS. Une fois le système conçu, nous avons évalué ses performances avec une application teste. Cette évaluation nous a donné des résultats encourageants.

Enfin, nous avons étudié les différentes failles de notre système.

# Contribution

Étant donné les objectifs que nous nous sommes fixés pour ce projet, les principales contributions de notre étude peuvent être résumées comme suit :

L'étude comparative des techniques de gestion de projet a permis de classifier différent technique de gestion inférence de proposer une technique adéquate à l'entreprise (l'équipe de développement et au projet).

L'étude du marché des différents logiciels de gestion de projet nous a permis de choisir un logiciel qui s'adapte le mieux au besoin de l'entreprise en terme de faciliter d'utilisation.

Aussi la mise en pratique de la méthode agile dans un projet concret a permis de voir concrètement quel est l'apport de celle-ci dans le développement des projets de l'entreprise.

# Perspectives et travaux futurs

Bien que les méthodes agiles soient applicables au développement de service à petit granularité, elles présentent certains inconvénients Globalement, plus d'un responsable informatique tique sur la légèreté de ces méthodes en matière de formalisation et de documentation. Il est donc intéressent d'étudier cette méthode dans des projets plus grands, ce qui va certainement commencé à apparaitre avec l'expansion de l'entreprise.

# Appréciation

Ce rapport a été rédigé à l'issu de mon stage de fin d'étude qui s'est déroulé au sien de l'entreprise Nomalys.

Ce stage a été pour moins une très bonne expérience tant sur le plan professionnel que personnel. En effet ce stage m'a permis de me former et surtout de mettre en pratique les toutes dernières technologies en vogue dans le monde .net, mais aussi une nouvelle approche de développement grâce aux méthodes agiles.

J'ai également pu découvrir et utiliser des outils de gestion de projet comme Redmine tout aussi puissant qu'indispensable de mon avis pour développer des applications aujourd'hui.

Enfin le fait d'avoir effectué plusieurs projets divers au niveau des fonctionnalités et des buts demandés a été un point intéressant et très constructif.

J'ai également découvert de l'intérieur de monde des startups vu que j'ai effectué mon stage dans l'incubateur de l'INT qui regroupe plus de 10 startups.

Ce stage de fin d'études a donc été très enrichissant sur le plan technique mais, aussi sur le plan relationnel managérial. J'en tire un bilan très positif.

# Référence

# Références Webographiques

**Annexe**

# Innovation Nomalys

3 niveaux d'innovation

# 1 -généralité

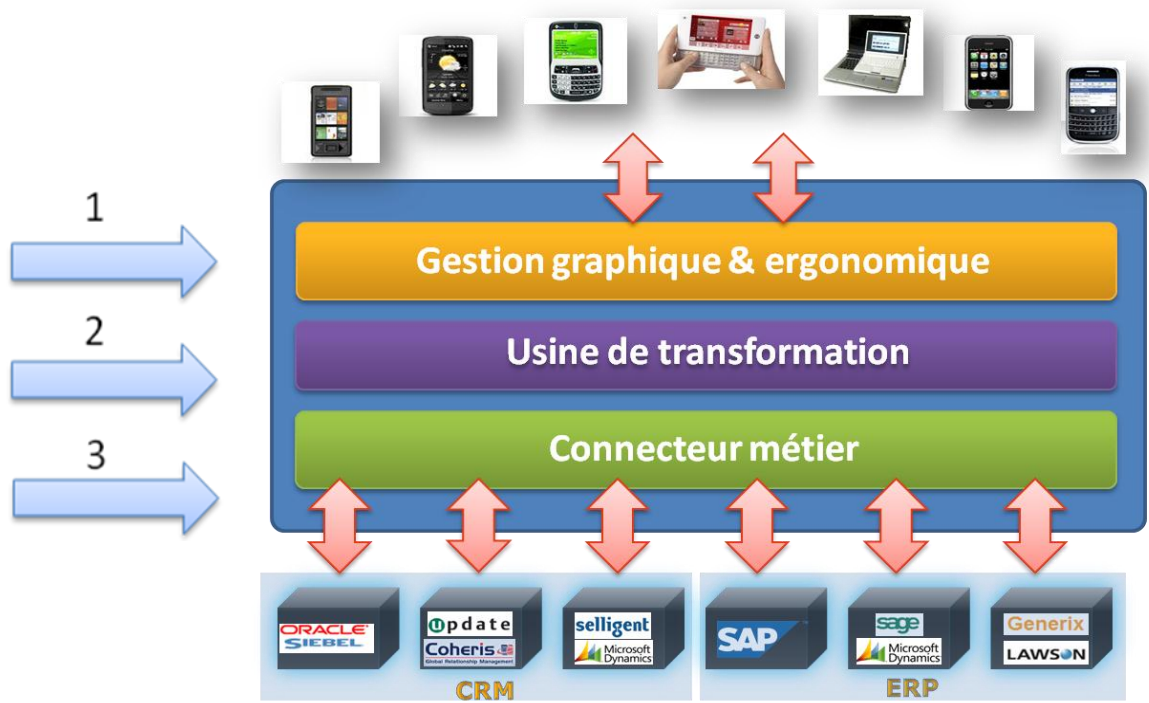

L'innovation de Nomalys est sur les 3 niveaux.

- -1- Gestion graphique
    - IHM totalement en rupture d'usage
- -2- Usine de transformation
    - Permet d'avoir une IHM nouvelle, un moteur de navigation unique pour tous les clients, une recherche type Google, une gestion d'alertes automatiques
- -3- Connecteur Métier
    - Est complètement indépendant du produit

# 2-Innovation IHM

a) Ce qui existe et que nous ne voulons pas

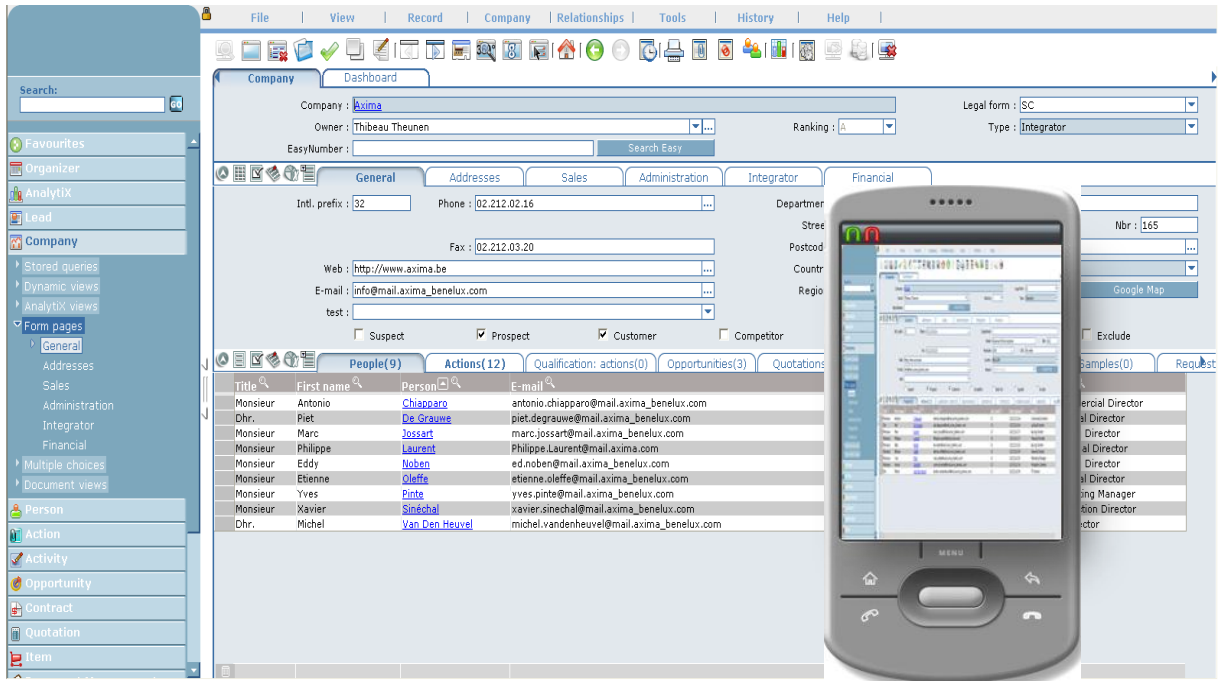

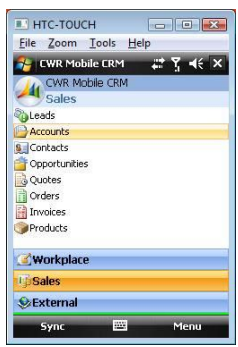

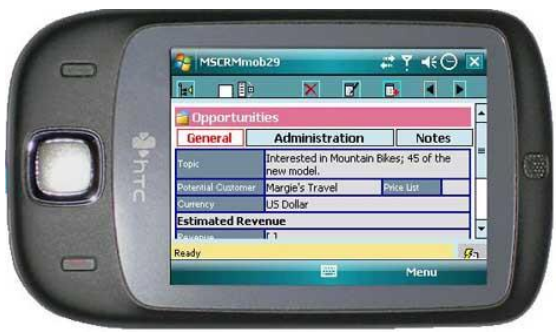

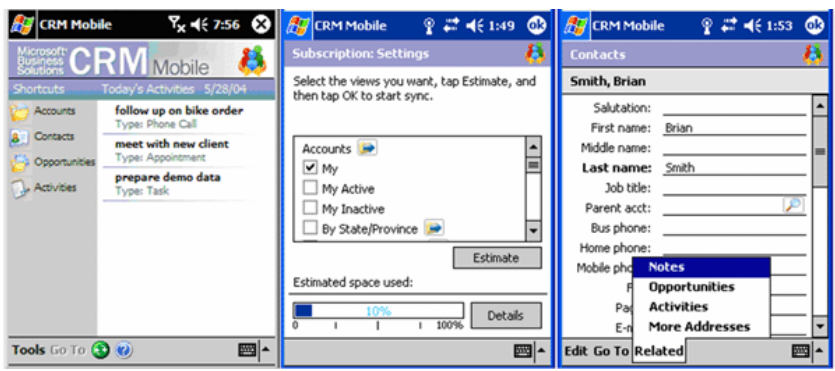

## Fonctionne pour le back office

Les produits existants adaptent l'ergonomie du produit initial au terminal cible.
Le but est, entre autre, de ne pas modifier les habitudes de l'utilisateur.
Cela peut fonctionner dans les métiers du back office où les utilisateurs sont OBLIGÉS d'utiliser leur PC et leurs terminaux afin de :
- Faire un relevé de compteur,
- entrer un facture,
- faire un relevé de linéaire,
- prendre une commande,
- mettre à jour un stock …

## Pas pour le front office

Les utilisateurs « front Office » sont beaucoup plus exigeants vis-à-vis de leurs produits.
Pour faire en sorte que les utilisateurs front office utilisent leurs terminaux mobiles, peu de manière coercitive, ont montré leur efficacité.

Si l'interface n'est pas belle, ergonomique, sexy, efficace. Ils n'utilisent pas.
Si le produit n'a pas une réelle valeur ajoutée autre que le flicage : ils n'utilisent pas.

Bien que très efficace pour les métiers du back Office, cette ancienne méthode d'adaptation des environnements PC vers des Smartphones ne fonctionne pas et n'a jamais fonctionné sur la population visée par Nomalys.

| b) Ce que nous proposons |

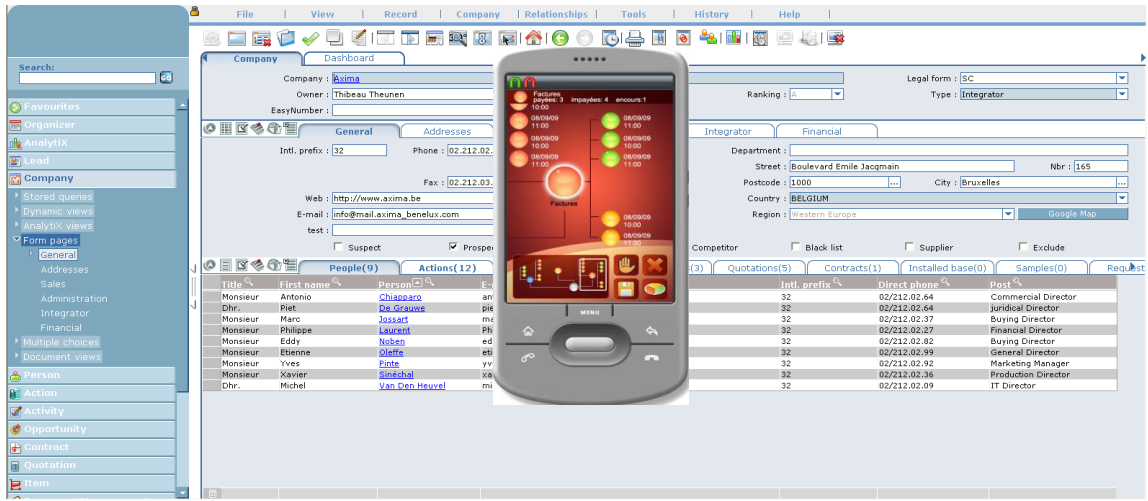

## Rupture d'usage

Nomalys propose une vraie rupture d'usage pour ces utilisateurs pressés et exigeants.
Nous ne nous inspirons pas des interfaces classiques des PC.
Notre interface est inspirée du monde des Smartphones et des règles d'ergonomie du jeu vidéo.

## Choix d'habillage

Nous proposons un large choix d'habillage de l'écran que l'utilisateur change suivant son humeur ☺
Un des habillages varie automatiquement suivant le type de donnée visualisée, un autre varie suivant l'heure du jour.

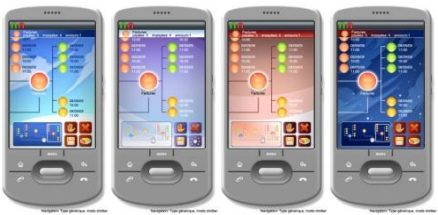

## On navigue par choix

La navigation est intuitive et on la veut agréable. L'utilisateur n'est plus contraint de naviguer dans ses données et cela devient un choix.

## Interface unique

Le moteur de navigation est identique pour tous les clients.
Il est donc indépendant du client, du type de données géré, du volume de données.

## D'un seul doigt

L'interface graphique a été pensée pour pouvoir être utilisable avec le pouce.
L'utilisateur peut choisir entre une interface pour droitier ou gaucher.

### c) Les innovations

## Le look

Nous faisons en sorte que l'utilisateur soit fier et heureux de se connecter à son application métier depuis son Smartphone

## Vertical/horizontale

Nous offrons à l'utilisateur un autre axe de visibilité et navigation au sein des données en adaptant le mode de visualisation suivant l'inclinaison de l'appareil. Cette adaptation ne restreint à faire une rotation de 90° de l'ensemble mais à offrir des fonctionnalités adaptées au mode vertical et au mode horizontal

## Accéssibilité

Nous offrons la possibilité au plus grand nombre de prendre du plaisir à utiliser Nomalys. Cela passe par un :
- Réglage du contraste,
- Choix des couleurs pour qu'elles soient compatibles mal voyants et daltoniens,
- Modification de la taille de police des caractères,
- Zones de saisie intelligentes acceptant l'imprécision due aux tremblements ou à un manque de coordination,
- Gestion des droitiers et gauchers,
- Choix de l'habillage graphique.

## Head Up

Emprunté aux techniques du jeu vidéo et à l'armée, Nomalys introduit les ingrédients de type Head-Up sur Smartphone.
Donner la possibilité à l'utilisateur d'avoir en même temps ses informations de navigation et les informations qui lui ont été remontées ceci dans le cadre d'une utilisation performante et intuitive.

## 3. Traitements

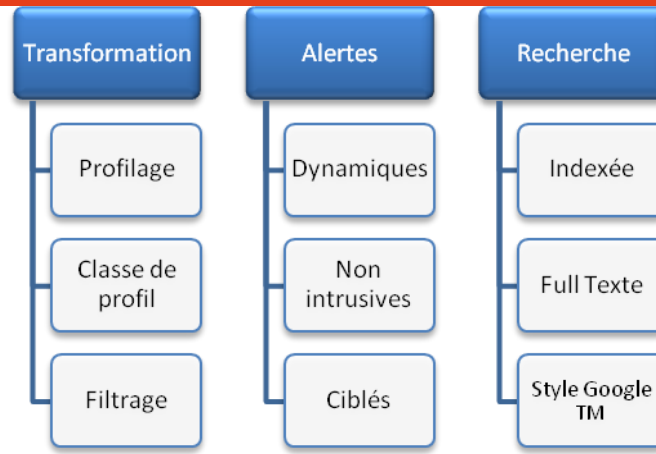

La couche de traitement, entre l'extraction des données et le Smartphone, possède le plus grand nombre d'innovations.

### c) Transformation.

Le but de la brique transformation est de choisir automatiquement pour l'utilisateur les informations importantes du dossier qu'il demande.

Les informations remontées vers le Smartphone sont remontées de manière dynamique et sont différentes suivant la personne et le contexte.

Le choix des informations se fait entre entre autre suivant le profil de l'utilisateur, profil qui est auto apprenant à l'aide des méthodes telles que l'approche dynamique ou par apprentissage.

Nomalys choisi également les informations suivant la classe de profil de l'utilisateur via la méthode dynamique.

La mise en œuvre du filtrage est la conjonction du profilage, de la classe de profil ainsi que les méthodes de filtrage telles que collaboratif, adaptatif ou actif.

### d) Alertes

Des alertes non intrusives sont poussées vers les Smartphones des utilisateurs. Ces alertes sont identifiées de manière automatique sur les données.

Ex : une facture vient d'être réglée, une commande vient d'être enregistrée, une réclamation urgente et grave vient d'être postée via la cellule SAV par un de mes clients.

### e) Recherche

Il est difficilement envisageable de faire une recherche champs à champs (QBE – Query by example) sur un médium de type Smartphone.

Nomalys Indexe les bases de données et permet de faire une recherche approximative sur l'ensemble des données depuis un seul champ.

Cette recherche s'apparente aux recherches existantes style Google TM.

## 4.Produit à interfacer

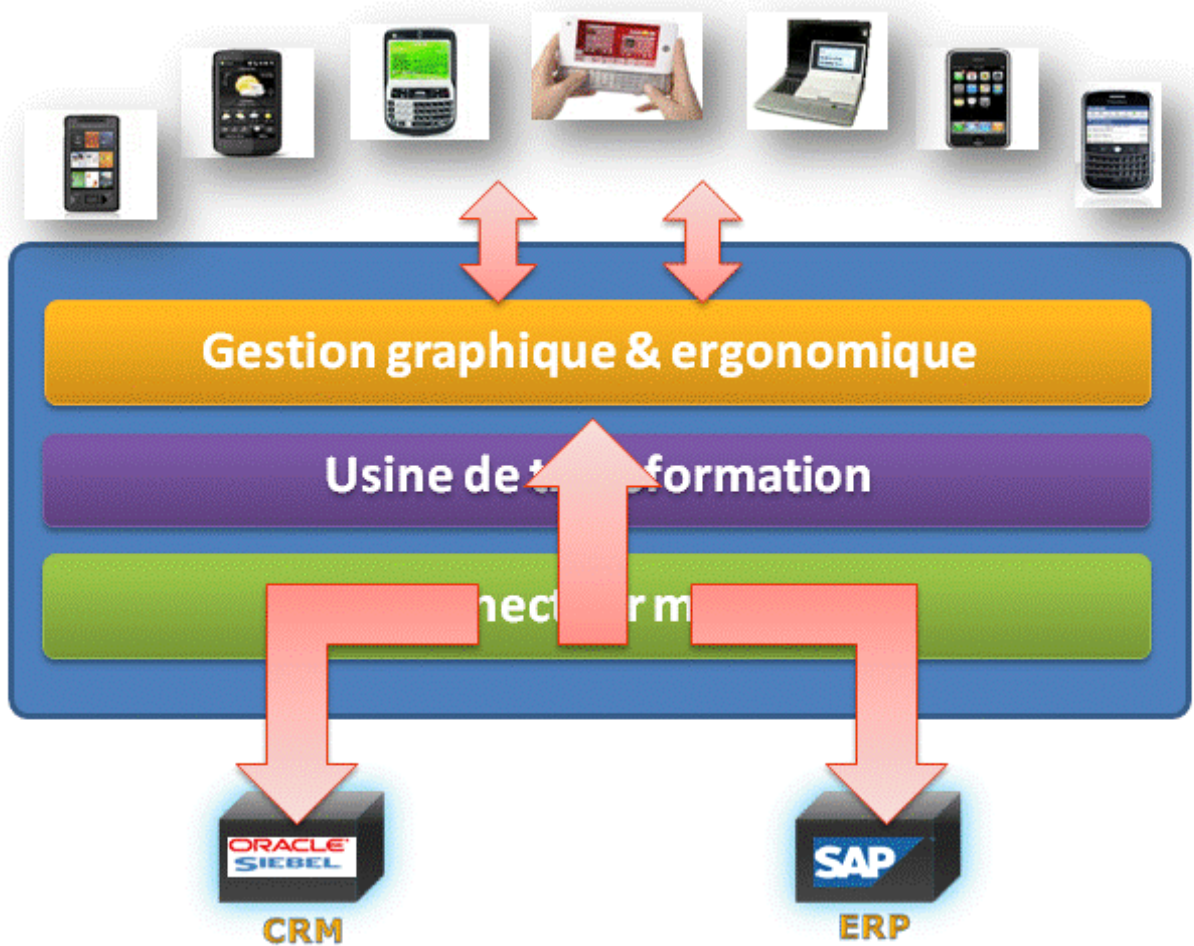

La couche Connecteur qui n'est pas lié à un produit en particulier permet de pouvoir, lors d'une requête, rechercher dans plusieurs produits / base de données, transformer l'information via la couche usine de transformation et pousser vers l'utilisateur un concentré d'informations pertinentes.